\documentclass[conference]{IEEEtran} 
\pdfoutput=1

\usepackage[pdftex]{graphicx}
\DeclareGraphicsExtensions{.pdf,.jpeg,.png}

\usepackage[cmex10]{amsmath}
\usepackage{amsfonts, amssymb}
\usepackage{amssymb} 
\usepackage{cite} 
\usepackage{hyperref}
\IEEEoverridecommandlockouts

\begin{document} 
\newcommand{\wt}[2]{W_{#1}[#2]}
\newcommand{\wh}[2]{\hat{W}_{#1}[#2]}
\newcommand{\ws}[3]{\textbf{W}_{#1;#2}^{#3}} 
\newcommand{\wsh}[3]{\hat{\textbf{W}}_{#1;#2}^{#3}}
\newcommand{\Nob}[3]{N_{ob,A_{#1}}^{(#2,#3)}}
\newcommand{\NobE}[3]{\overline{N}_{ob,A_{#1}}^{(#2,#3)}}
\newcommand{\NobTwo}[2]{N_{ob,A_{#1}}^{(#2)}}
\newcommand{\NobETwo}[2]{\overline{N}_{ob,A_{#1}}^{(#2)}}
\newcommand{\Pob}[3]{P_{ob,A_{#1}}^{(#2,#3)}}
\newcommand{\pr}{\text{Pr}} 
\newcommand{\xstar}{n}
\newcommand{\ystar}{q} 
\newcommand{\tx}{TX} 
\newcommand{\rx}{RX} 
\newcommand{\xiN}[1]{\xi_{#1}} 
\title{Analysis and Design of Two-Hop Diffusion-Based Molecular Communication Networks}
\author{ 
\IEEEauthorblockN{Arman Ahmadzadeh\IEEEauthorrefmark{1}, Adam Noel\IEEEauthorrefmark{2}, and Robert Schober\IEEEauthorrefmark{1}} 
\IEEEauthorblockA{\IEEEauthorrefmark{1}University of Erlangen-Nuremberg, Germany} 
\IEEEauthorblockA{\IEEEauthorrefmark{2}University of British Columbia, Canada}
\thanks{This work was supported in part by the AvH Professorship Program of the Alexander von Humboldt Foundation.} 
}
\maketitle

\begin{abstract}
In this paper, we consider a two-hop molecular communication network consisting of one nanotransmitter, one nanoreceiver, and one nanotransceiver acting as a relay. We consider two different schemes for relaying to improve the range of diffusion-based molecular communication. In the first scheme, two different types of messenger molecules are utilized at the relay node for transmission and detection. In the second scheme, we assume that there is only one type of molecule available to be used as an information carrier. We identify self-interference as the performance-limiting effect for the second relaying scheme. Self-interference occurs when the relay must detect the same type of molecule that it also emits. Furthermore, we consider two relaying modes analogous to those used in wireless communication systems, i.e., full-duplex and half-duplex. In particular, while our main focus is on full-duplex relaying, half-duplex relaying is employed as a means to mitigate self-interference. In addition, we propose the adaptation of the decision threshold as an effective mechanism to mitigate self-interference at the relay for full-duplex transmission. We derive closed-form expressions for the expected error probability of the network for both considered relaying schemes.             
\end{abstract}

\IEEEpeerreviewmaketitle
\section{Introduction}
Exchanging information via molecules, i.e., molecular communication (MC), is used by nature for communication among biological entities ranging from cells to organs \cite{AlbertsBook}. The biocompatibility of MC makes it an attractive candidate for enabling communication between so-called nanomachines, i.e., small-scale devices having functional units on the order of nanometers in size \cite{Nakano1}. The ultimate goal of integrating communication capabilities into nanomachines is to increase their functionality such that communities of nanomachines, so-called nanonetworks, can perform collaborative and challenging tasks in a distributed manner. It is envisioned that nanonetworks have biomedical, environmental, and industrial applications \cite{NakanoB}. 

In diffusion-based MC, the molecules released by the transmitter into a fluid environment diffuse in all directions without any additional infrastructure and some of them reach the receiver. However, diffusion-based MC poses unique challenges that are not commonly found in traditional communication networks, and which have to be carefully addressed for the development of such networks. One of the challenges of diffusive MC is that the propagation time increases with the square of the distance. If an intended receiver is far away, then using a single transmitter may be impractical. One approach from conventional wireless communications that could be adapted for MC to aid communication with distant receivers is the use of relays.

Relays can potentially improve the reliability of a communication link. In fact, the relaying of information also plays an important role in communication among biological systems. For example, in typical communication between cells, a \textit{signaling cell} produces a particular type of signal molecule that is detected by a \textit{target cell}. The target cell possesses \textit{receptor proteins} that recognize and respond specifically to the signal molecule. When a signal molecule is detected by a cell-surface receptor, this information is relayed into the interior of the target cell via a set of \textit{intracellular signaling molecules}, which act in sequence and ultimately change the behavior of the target cell; see \cite[Ch. 16]{AlbertsBook}. 

Several works have recently addressed multihop communication among nanomachines. In \cite{Einolghozati1} and \cite{Einolghozati2}, a diffusion-based multihop network between bacteria colonies was analyzed, where each node of the network was formed by a population of bacteria. In \cite{Nakano2} and \cite{Nakano3}, the design and analysis of repeater cells in Calcium junction channels were investigated. In \cite{Unluturk1}, the rate-delay trade-off of a three-node nanonetwork was analyzed for a specific messenger molecule, polyethylene, and network coding at the relay node. The use of bacteria and virus particles as information carriers in a multihop network was proposed in \cite{Balasubramaniam1} and \cite{Walsh1}, respectively.

In this paper, we consider a two-hop MC network and investigate two different relaying schemes. In the first scheme, we assume that each hop uses a different type of messenger molecule. In the second scheme, we assume that the same type of messenger molecule is used in both hops. We identify self-interference at the relay as the effect that limits the performance of the second relaying scheme with full-duplex transmission. Specifically, self-interference occurs when the relay must emit and detect the same type of molecule. We propose two approaches for self-interference mitigation. In the first approach, in every bit interval, the detection threshold at the relay is adjusted based on the knowledge of all previously-detected information bits. In the second approach, half-duplex relaying is adopted instead of full-duplex relaying. We derive closed-form analytical expressions for the expected error probability of the considered two-hop network for both relaying schemes based on the error rate analysis of a single link reported in \cite{NoelJ1}, \cite{NoelJ2}. 

The rest of this paper is organized as follows. In Section \ref{Sec.SysModandPre}, we introduce the system model and preliminaries of error rate analysis. In Section \ref{Sec.Two-hopNetPerAna}, we evaluate the expected error probability of the two-hop network for both of the considered relaying schemes. Numerical results are given in Section \ref{Sec.NumRes}, and conclusions are drawn in Section \ref{Sec.Con}. 

\section{SYSTEM MODEL AND PRELIMINARIES} 
\label{Sec.SysModandPre} 
\subsection{System Model}
\begin{figure}[!t]
	\centering
	\includegraphics[scale = 0.6]{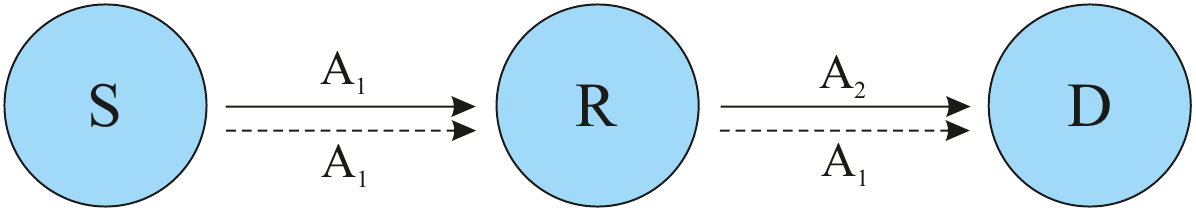}\vspace*{-3mm}
	\caption{System model of a two-hop MC network, where the first and the second relaying schemes are illustrated with solid and dashed arrows, respectively.}
	\label{Fig.SysMod}
\end{figure} 
In this paper, we use the terms ``nanomachine'' and ``node'' interchangeably to refer to the devices in the network, as the term ``node" is commonly used in the relaying literature. We assume that a source (S) node and a destination (D) node are placed at locations $(0,0,0)$ and $(x_{D},0,0)$ of a 3-dimensional space, respectively. The relay (R) node is placed in the middle between node S and node D along the $x$-axis, cf. Fig. \ref{Fig.SysMod}. We assume that node D and node R are spherical in shape with fixed volumes (and radii) $V_{D}$ $(r_{D})$ and $V_{R}$ $(r_{R})$, respectively, and that they are passive observers such that molecules can diffuse through them. 

We consider two different relaying schemes. For the first scheme, we assume that there are two different types of messenger molecules, $A_{1}$ and $A_{2}$, and that relay R can detect type $A_{1}$ molecules, which are released by node S, and emits type $A_{2}$ molecules, which are detected by node D. For the second scheme, we assume that relay R uses the same type of molecules, $A_{1}$, for transmission \emph{and} detection. The number of released molecules of type $A_{f}, f \in \{ 1,2\}$, is denoted as $N_{A_{f}}$, and the concentration of type $A_{f}$ molecules at the point defined by vector $\vec{r}$ at time $t$ in molecule $\cdot$ $\text{m}^{-3}$ is denoted by $C_{A_{f}}( \vec{r}, t )$. We assume that the movements of different molecules are independent. 

Furthermore, we assume that the information that is sent from node S to node D is encoded into a binary sequence of length $K$, $\textbf{W}_{S} = \{ \wt{S}{1}, \wt{S}{2},..., \wt{S}{K} \}$. Here, $\wt{S}{j}$ is the bit transmitted by node S in the $j$th bit interval with $\pr(\wt{S}{j}=1)=P_{1}$, and $\pr(\wt{S}{j}=0)=P_{0}=1-P_{1}$, where $\pr(\cdot)$ denotes probability. The information bits transmitted and detected by relay R in the $j$th bit interval are denoted by $\wt{R}{j}$ and $\wh{R}{j}$, respectively. The information bit detected at node D in the $j$th bit interval is denoted by $\wh{D}{j}$. In the following, we use a common notation to denote a subsequence transmitted (or detected) by node $h$, $h \in \{S,R,D\}$. In particular, we write $\ws{h}{a}{b} = \{ \wt{h}{a},...,\wt{h}{b} \}$ and $ \wsh{h}{a}{b} = \{ \wh{h}{a},...,\wh{h}{b} \}$ for transmission and detection, respectively. We adopt ON/OFF keying for modulation and a fixed bit interval duration of $T_{B}$ seconds. Node S releases $N_{A_{1}}$ molecules at the beginning of the bit interval to convey information bit ``1'', and no molecules to convey information bit ``0''. This is a commonly-used modulation scheme in the MC literature; see e.g. \cite{Nakano4, NoelJ3, Atakan1, Llatser1}. Furthermore, we adopt the decode-and-forward (DF) relaying protocol, where the relay first decodes the received message, and then re-encodes the detected message for re-transmission. 

We consider two protocols for two-hop transmission, namely full-duplex and half-duplex. For full-duplex transmission, reception and transmission occur simultaneously at the relay node, i.e., in each bit interval, relay R detects the information transmitted by node S, and sends the information bit detected in the previous bit interval to node D. For half-duplex transmission, the relay performs detection and reception separately, i.e., in one bit interval, relay R detects the information transmitted by node S, and in the next bit interval, relay R transmits the detected information bit to node D. For transmitting $L$ bits, node S sends $\ws{S}{1}{K}$, where $K=L$ and $K=2L$ for full-duplex and half-duplex relaying, respectively. For half-duplex relaying, node S transmits the $L$ information bits in the odd bit intervals, i.e., $\wt{S}{2i-1}, i \in \{1,...,L\}$. In even bit intervals, node S is silent, i.e., $\wt{S}{2i}=0, i \in \{1,...,L\}$, and node R does not detect.
 
\subsection{Preliminaries} 
In the following, we consider a general communication between a transmitting node $\xstar \in \{S,R\}$ and a receiving node $\ystar \in \{R,D\}$, $\xstar \neq \ystar$, and review the corresponding error rate analysis provided in \cite{NoelJ1}, \cite{NoelJ2}. We highlight the parts that are utilized in the analysis of the two-hop network. In the following, $A$ is the type of molecule released by node $\xstar$ and detected at node $\ystar$. 

The independent diffusion of molecules through the environment can be described by Fick's second law as \cite[Eq. (3)]{NoelJ1} 
\begin{equation}
	\label{Eq. Fick's Second Law} 
	\frac{\partial C_{A}(\vec{r},t)}{\partial t} = D_{A} \nabla^2 C_{A}(\vec{r},t),
\end{equation} 
where $D_{A}$ is the diffusion coefficient of $A$ molecules in $\frac{\text{m}^2}{\text{s}}$. Assuming that node $\xstar$ is an impulsive point source, and emits $N_{A}$ molecules at the point defined by vector $\vec{r}_{\xstar}$ into an infinite environment at time $t = 0$, the local concentration at the point defined by vector $\vec{r}$ and at time $t$ is given by \cite[Eq. (4)]{NoelJ1} 
\begin{equation}
	\label{Eq. Local Concentration} 
	C_{A}(\vec{r},t) = \frac{N_{A}}{(4 \pi D_{A} t)^{3/2}} \exp \left( -\frac{\vert \vec{r} - \vec{r}_{\xstar}\vert^2}{4D_{A}} \right).
\end{equation} 
It is shown in \cite{NoelPro1} that the number of molecules observed within the volume of node $\ystar$, $V_{\ystar}$, at time $t$ due to one emission of $N_{A}$ molecules at $\vec{r}_{\xstar}$ at $t=0$, $\Nob{}{\xstar}{\ystar}(t)$, can be accurately approximated as a Poisson random variable with time-varying mean given by
\begin{equation}
	\label{Eq. NobE} 
	\NobE{}{\xstar}{\ystar}(t) = C_{A}(\vec{r}_{\ystar},t)V_{\ystar},
\end{equation}
where $\vec{r}_{\ystar}$ is the vector from the origin to the center of node $\ystar$, and we used the uniform concentration assumption, i.e., we assumed that node $\ystar$ is a point observer or that the concentration throughout its volume is uniform and equal to that at its center; see \cite{NoelPro1} for the conditions necessary for the validity of this assumption. The probability of observing a given $A$ molecule, emitted by node $\xstar$ at $t=0$, inside $V_{\ystar}$ at time $t$, $\Pob{}{\xstar}{\ystar}(t)$, is given by (\ref{Eq. NobE}) when setting $N_{A}=1$, i.e.,
\begin{equation}
	\label{Eq. Pob} 
	\Pob{}{\xstar}{\ystar}(t) = \frac{V_{\ystar}}{(4 \pi D_{A} t)^{3/2}} \exp \left( -\frac{\vert \vec{r}_{\ystar} -\vec{r}_{\xstar}\vert^2}{4D_{A}} \right). 
\end{equation} 
For detection, we adopt a family of receivers introduced in \cite{NoelJ2}, the so-called weighted sum detectors, where the receiving node takes samples equally spaced in time within a single bit interval, adds up the individual samples with a certain weight assigned to each sample, and then compares the sum with a decision threshold. For simplicity, we assume equal weights for all samples. The decision of the weighted sum in the $j$th bit interval is given by \cite[Eq. (37)]{NoelJ2} 
\begin{equation}
	\label{Eq.Reception} 
	\wh{\ystar}{j} = \begin{cases} 
	1 &\mbox{if } \sum_{m=1}^{M} \Nob{}{\xstar}{\ystar}(t(j,m)) \geq \xi_{\ystar}, \\
	0 &\mbox{otherwise,} 
			\end{cases}
\end{equation} 
where $\xi_{\ystar}$ is the binary detection threshold of node $\ystar$, and we assume that node $\ystar$ takes $M$ samples in each bit interval. The sampling time of the $m$th sample in the $j$th bit interval is $t(j,m) = (j-1)T_{B} + t_{m}$, where $t_{m}=mt_{0}$ and $t_{0}$ is the time between two successive samples. $\Nob{}{\xstar}{\ystar}(t(j,m))$ is a Poisson random variable with mean $\NobE{}{\xstar}{\ystar}(t(j,m))$ for any individual sample. Thus, the sum of all samples in the $j$th bit interval, $\Nob{}{\xstar}{\ystar}[j]=\sum_{m=1}^{M} \Nob{}{\xstar}{\ystar}(t(j,m))$, is also a Poisson random variable whose mean is the sum of the means of the individual samples, i.e., $\NobE{}{\xstar}{\ystar}[j] = \sum_{m=1}^{M}\NobE{}{\xstar}{\ystar}(t(j,m))$. Due to the independent movement of molecules, node $\ystar$ observes molecules that were emitted at the start of the current or any prior bit interval. As a result, the number of molecules observed within $V_{\ystar}$ in the $j$th bit interval due to the transmission of sequence $\ws{\xstar}{1}{j}$, $\Nob{}{\xstar}{\ystar}[j]$, is also a Poisson random variable with mean 
\begin{equation}
	\label{Eq. NobE Cumulative} 
	\NobE{}{\xstar}{\ystar}[j] = N_{A} \sum_{i=1}^{j} \wt{\xstar}{i} \sum_{m=1}^{M} \Pob{}{\xstar}{\ystar}((j-i)T_{B}+t_{m}).
\end{equation} 

Given $\ws{\xstar}{1}{j-1}$ and assuming that there is no \textit{a priori} knowledge about $\wt{\xstar}{j}$, the probability of error in the $j$th bit interval can be written as 
\begin{align} 
	\label{Eq. ExpErrorOneHop}
	P_{e_{1}}[j|\ws{\xstar}{1}{j-1}] &= P_{1}\pr(\Nob{}{\xstar}{\ystar}[j] < \xi_{\ystar} | \wt{\xstar}{j}=1,\ws{\xstar}{1}{j-1}) \nonumber \\ 
	& + P_{0}\pr(\Nob{}{\xstar}{\ystar}[j] \geq \xi_{\ystar} | \wt{\xstar}{j}=0,\ws{\xstar}{1}{j-1}),
\end{align} 
where the cumulative distribution function (CDF) of the weighted sum in the $j$th bit interval is given by \cite[Eq. (38)]{NoelJ2} 
\begin{align}
	\label{Eq. CDF} 
	\pr \left( \Nob{}{\xstar}{\ystar}[j] < \xi_{\ystar} | \ws{\xstar}{1}{j} \right) &= \exp (-\NobE{}{\xstar}{\ystar}[j]) \nonumber \\
     & \times \sum_{\omega=0}^{\xi_{\ystar}-1} \frac{\left( \NobE{}{\xstar}{\ystar}[j] \right)^{\omega}}{\omega!}.
\end{align}
The average error probability in the $j$th bit interval, $\overline{P}_{e_{1}}[j]$, is obtained by averaging $P_{e_{1}}[j|\ws{\xstar}{1}{j-1}]$ over all possible realizations of $\ws{\xstar}{1}{j-1}$.   
\section{TWO-HOP NETWORK PERFORMANCE ANALYSIS}
\label{Sec.Two-hopNetPerAna} 
In this section, we evaluate the expected error probability of the two-hop network for the two proposed relaying schemes, i.e., for different and identical types of molecule in both hops, respectively.
\subsection{Different Types of Molecules in Each Hop} 
Node S emits type $A_{1}$ molecules, which have diffusion coefficient $D_{A_{1}}$ and can be detected by relay node R. The relay emits type $A_{2}$ molecules having diffusion coefficient $D_{A_{2}}$ for forwarding the detected message to node D. Node S and node R release $N_{A_{1}}$ and $N_{A_{2}}$ molecules to transmit bit ``1'' at the beginning of a bit interval, respectively.

Since molecules of different types do not interfere with each other, we only consider full-duplex relaying in this case. The nodes communicate as follows. At the beginning of the $j$th bit interval, node S transmits information bit $\wt{S}{j}$, and node R transmits concurrently the information bit detected in the previous bit interval, $\wt{R}{j} = \wh{R}{j-1}$. At the end of the $j$th bit interval, node R and node D make decisions on the respective received signals. Thus, node D receives the $j$th bit with one bit interval delay. The total duration of transmission for a sequence of length $L$ is $(L+1)T_{B}$. 

In the two-hop communication link, for binary modulation, an error occurs if the detection is erroneous in either the first hop or the second hop. Given $\wt{S}{j}$, an error occurs in the $(j+1)$th bit interval if $\wh{R}{j} \neq \wt{S}{j}$ and $\wh{D}{j+1} = \wt{R}{j+1}$, or if $\wh{R}{j} = \wt{S}{j}$ and $\wh{D}{j+1} \neq \wt{R}{j+1}$. Thus, the error probability of the $j$th bit can be written as 
\begin{align}
	\label{Eq. ErrorExpCurBit} 
	P_{e_{2}}[j] &= \pr( \wt{S}{j} \neq \wh{R}{j}) \times \pr(\wt{R}{j+1} = \wh{D}{j+1}) \nonumber \\
	& + \pr( \wt{S}{j} = \wh{R}{j}) \times \pr(\wt{R}{j+1} \neq \wh{D}{j+1}).
\end{align}        
Let us assume that $\ws{S}{1}{j-1}$ is given, and there is no \textit{a priori} knowledge about $\wt{S}{j}$ and $\wh{R}{j} = \wt{R}{j+1}$. Then, the error probability of the $j$th bit is given by 
\begin{align}
	\label{Eq. ErrorExpSeq} 
	  P_{e_{2}}[j|\ws{S}{1}{j-1}] &= P_{1}\pr (\Nob{1}{S}{R}[j] < \xiN{R} | \wt{S}{j}=1, \ws{S}{1}{j-1}) \times \nonumber \\
	 & \pr(\Nob{2}{R}{D}[j+1] < \xiN{D} | \wt{R}{j+1} = 0, \wsh{R}{1}{j-1}) \nonumber \\ 
	& + P_{0}\pr (\Nob{1}{S}{R}[j] \geq \xiN{R} | \wt{S}{j}=0, \ws{S}{1}{j-1}) \times \nonumber \\
	 & \pr(\Nob{2}{R}{D}[j+1] \geq \xiN{D} | \wt{R}{j+1} = 1, \wsh{R}{1}{j-1}) \nonumber \\ 
	& + P_{1}\pr (\Nob{1}{S}{R}[j] \geq \xiN{R} | \wt{S}{j}=1, \ws{S}{1}{j-1}) \times \nonumber \\  
	 & \pr(\Nob{2}{R}{D}[j+1] < \xiN{D} | \wt{R}{j+1} = 1, \wsh{R}{1}{j-1}) \nonumber \\ 
	& + P_{0}\pr (\Nob{1}{S}{R}[j] < \xiN{R} | \wt{S}{j}=0, \ws{S}{1}{j-1}) \times  \nonumber \\
	 & \pr(\Nob{2}{R}{D}[j+1] \geq \xiN{D} | \wt{R}{j+1} = 0, \wsh{R}{1}{j-1}).
\end{align} 
For a given $\ws{S}{1}{j-1}$, there are $2^{(j-1)}$ different possible realizations of $\wsh{R}{1}{j-1}$. However, in (\ref{Eq. ErrorExpSeq}), to keep the complexity of evaluation low, we consider only one realization of $\wsh{R}{1}{j-1}$ which leads to an approximation. In particular, this realization of $\wsh{R}{1}{j-1}$ is obtained via a biased coin toss. To this end, we model the detected bits in $\wsh{R}{1}{j-1}$ , i.e., $\wh{R}{i}, i \in \{1,2,...,j-1 \}$, as $\wh{R}{i} = |\lambda - \wt{S}{i} |$, where $\lambda \in \{0,1 \}$ is the outcome of the coin toss with $\pr(\lambda = 1) = P_{e_{1}}[i|\ws{S}{1}{i-1}]$ and $\pr(\lambda = 0) = 1-P_{e_{1}}[i|\ws{S}{1}{i-1}]$. Our simulation results in Section \ref{Sec.NumRes} confirm the accuracy of this approximation.
 
\subsection{Same Type of Molecule in Each Hop} 
We now consider a two-hop network where the same type of molecule is employed in both hops. In this case, node S releases molecules of type $A_{1}$, which are detected by relay node R. Node R utilizes the same type of molecule, $A_{1}$, for forwarding the detected message to node D. We first consider full-duplex transmission. When the relay uses the same type of molecule for detection and transmission, some of the molecules released by the relay at the beginning of a bit interval stay nearby and are observed during the bit interval and in subsequent bit intervals inside $V_{R}$. This effect, due to the random walks of the molecules, causes \textit{self-interference}. Let us consider a short example to clarify the occurrence of self-interference. Let us assume that the information sequence emitted by node S is ``10'', i.e., $\wt{S}{j}=1, \wt{S}{j+1}=0, \text{and } \wt{S}{m}=0, m<j$, and that no error occurs in the transmission of $\wt{S}{j}$ to node R, i.e., $\wh{R}{j}=\wt{S}{j}=1$. At the beginning of the $(j+1)$th bit interval, node R releases $N_{A_{1}}$ molecules to forward the detected message to node D, i.e., $\wt{R}{j+1}=1$. Due to the random walk of the molecules, some of the molecules released by the relay node may be observed within its own volume, $V_{R}$, at the time of sampling for detection of $\wt{S}{j+1}$. This causes self-interference and may lead to an erroneous decision for $\wh{R}{j+1}$.

In the following, we propose two approaches to mitigate the self-interference: 1) Employing an adaptive decision threshold at the relay, and 2) employing half-duplex relaying instead of full-duplex relaying. 

\textit{1) Adaptive Decision Threshold:} In the first approach, the relay adjusts its decision thresholds in each bit interval based on its detected information bits in all previous bit intervals. The adaptive decision threshold of the relay in the $j$th bit interval, $\xi_{R,Adp}[j]$, consists of two parts. The first part is a fixed threshold, $\xi$, and the second part, $\xi_{Exp}[j]$, changes adaptively based on the number of molecules expected within $V_{R}$ due to the bits detected at the relay prior to the current bit interval, $\wsh{R}{1}{j-1}$, i.e., 
\begin{equation}
	\label{Eq. Adp Thresh1} 
		\xi_{R,Adp}[j]=\xi + \xi_{Exp}[j].
\end{equation}
 
For optimizing $\xi_{Exp}[j]$, we have to determine the probability of observing a given molecule transmitted by the relay node at $t=0$ within $V_{R}$ at time $t$. We denote this probability as $\Pob{1}{R}{R}(t)$. 
$\Pob{1}{R}{R}(t)$ may be considered as a special case of $\Pob{}{\xstar}{\ystar}(t)$ when $\xstar = \ystar$, i.e., $\vec{r}_{\ystar} = \vec{r}_{\xstar}$. However, in this case, the conditions necessary for the validity of the uniform concentration assumption do not hold \cite{NoelPro1}. Hence, we can not use (\ref{Eq. Pob}) to evaluate $\Pob{1}{R}{R}(t)$. The general form of $\Pob{}{\xstar}{\ystar}(t)$, if the uniform concentration assumption is not made, is given by \cite[Eq. (27)]{NoelPro1}. It can be shown that, by using l'H$\hat{\text{o}}$pital's rule, $\Pob{1}{R}{R}(t)$ in the limit of  $|\vec{r}_{\ystar} - \vec{r}_{\xstar}| \rightarrow 0$ can be written as 
\begin{equation}
	\label{Eq. Pselfobs} 
	\Pob{1}{R}{R}(t) = \text{erf}\left(  \frac{r_{R}}{2\sqrt[]{D_{A_{1}}t}} \right) - \frac{ r_{R} \exp{\left( \frac{{-r_{R}}^{2}}{4D_{A_{1}}t} \right)}}{\sqrt[]{\pi D_{A_{1}}t}}, 
\end{equation}
where $r_{R}$ is the radius of the relay node, and $\text{erf}(\cdot)$ denotes the error function as defined by \cite[Eq. 7.1.1]{abramowitz}. 

Thus, given $\wsh{R}{1}{j-1}$, the expected number of molecules observed within $V_{R}$ in the $j$th bit interval, $\NobE{1}{R}{R}[j]$, can be written as 
\begin{equation} 
	\label{Eq. ExpSelfOb} 
	\NobE{1}{R}{R}[j] = N_{A_{1}} \sum_{i=1}^{j-1} \wh{R}{i} \sum_{m=1}^{M} \Pob{1}{R}{R}((j-i)T_{B}+t_{m}), 
\end{equation} 
and the varying part of the adaptive decision threshold of the relay becomes $\xi_{Exp}[j] = \NobE{1}{R}{R}[j]$. The number of molecules observed inside $V_{R}$ in the $j$th bit interval when only the relay node transmits, $\Nob{1}{R}{R}[j]$, is a Poisson random variable with the mean given by (\ref{Eq. ExpSelfOb}). We define the complete received signal at the relay node in the $j$th bit interval as $\NobTwo{1}{R}[j]$, which is the sum of two signals, i.e., $\NobTwo{1}{R}[j] = \Nob{1}{S}{R}[j] + \Nob{1}{R}{R}[j]$. Since $\Nob{1}{S}{R}[j]$ and $\Nob{1}{R}{R}[j]$ are Poisson random variables with time-varying means, $\NobTwo{1}{R}[j]$ is also a Poisson random variable whose mean is the sum of the means of the individual variables, i.e., $\NobETwo{1}{R}[j] = \NobE{1}{S}{R}[j] + \NobE{1}{R}{R}[j] $. Analogously, the received signal at node D in the $j$th bit interval, $\NobTwo{1}{D}[j]$, is the sum of two Poisson random variables $\Nob{1}{S}{D}[j]$ and $\Nob{1}{R}{D}[j]$. Thus, $\NobTwo{1}{D}[j]$ is also a Poisson random variable with time-varying mean $\NobETwo{1}{D}[j] = \NobE{1}{S}{D}[j] + \NobE{1}{R}{D}[j]$. Finally, the expected error probability of the considered network can be evaluated via (\ref{Eq. ErrorExpSeq}), after substituting $\Nob{1}{S}{R}[j]$, $\Nob{2}{R}{D}[j]$, and $\xi_{R}$ with $\NobTwo{1}{R}[j]$, $\NobTwo{1}{D}[j]$, and $\xi_{R,Adp}[j]$, respectively, and considering that all conditional probabilities in (\ref{Eq. ErrorExpSeq}) have to be conditioned on both $\ws{S}{1}{j-1}$ and $\wsh{R}{1}{j-1}$. The detected bits in $\wsh{R}{1}{j-1}$, i.e., $\wh{R}{i}, i<j$, are modelled as $\wh{R}{i} = |\beta-\wt{S}{i}|$, where $\beta \in \{0,1\}$ is the outcome of a coin toss with $\pr(\beta=1)= P_{e_{1}}[i|\ws{S}{1}{i-1},\wsh{R}{1}{i-1}]$ and $\pr(\beta=0) = 1 - \pr(\beta =1)$. $P_{e_{1}}[i|\ws{S}{1}{i-1},\wsh{R}{1}{i-1}]$ can be evaluated via (\ref{Eq. ExpErrorOneHop}), after substituting $\Nob{}{\xstar}{\ystar}[i]$ and $\xi_{\ystar}$ with $\NobTwo{1}{R}[i]$ and $\xi_{R,Adp}[i]$, respectively. 

\textit{2) Half-Duplex Relaying:} In the second approach to mitigate self-interference, half-duplex relaying is adopted. In half-duplex relaying, reception and transmission at the relay occur in two consecutive bit intervals, giving the molecules released at the relay node time to leave $V_{R}$, so that they are less likely to interfere with the relay's decisions. 

For half-duplex relaying, the nodes communicate as follows. In odd bit intervals, S transmits and R receives, and in even bit intervals, R transmits and D receives. In other words, in the $(2j-1)$th bit interval, node S transmits the $j$th information bit, i.e., $\wt{S}{2j-1}$, which is detected by node R as $\wh{R}{2j-1}$, and in the $(2j)$th bit interval, node R transmits the $j$th bit detected in the previous bit interval, i.e., $\wt{R}{2j} = \wh{R}{2j-1}$. This bit is then detected at node D as $\wh{D}{2j}$. The total duration of transmission for a sequence of length $L$ is $2LT_{B}$.  

The expected error probability for half-duplex relaying can be evaluated via (\ref{Eq. ErrorExpSeq}), after substituting $\Nob{1}{S}{R}[j]$, $\Nob{2}{R}{D}$, $\wt{S}{j}$, and $\wt{R}{j+1}$ with $\NobTwo{1}{R}[2j-1]$, $\NobTwo{1}{D}[2j]$, $\wt{S}{2j-1}$, and $\wt{R}{2j}$, respectively, and considering that all conditional probabilities have to be conditioned on both $\ws{S}{1}{2j-1}$ and $\wsh{R}{1}{2j-1}$, where $\wt{S}{2i} = \wh{R}{2i}= 0 $ for $i \in \{1,2,...,(j-1) \}$.
\section{NUMERICAL RESULTS}
\label{Sec.NumRes} 
In this section, we present simulation and analytical results to evaluate the performance of the proposed relaying schemes. We adopted the particle-based stochastic simulator introduced in \cite{NoelJ1}. In our simulations, time is advanced in discrete steps $t_{0}$, i.e., the time between two consecutive samples, where in each time step molecules undergo random motion.
 
In order to focus on a comparison of the performance of the different relaying protocols, we keep the physical parameters of the relay and the receiver constant throughout this section. In particular, we assume that $ r_{R} = r_{D} = 45 \text{ nm}$, and both node R and node S release the same number of molecules. The only parameters that we vary are the decision threshold, the modulation bit interval, and the frequency of sampling. 

We consider a fluid environment with constant temperature and viscosity, and that the messenger molecules of all considered types have the same diffusion coefficient $D_{A_{1}} = D_{A_{2}} = 4.365 \times 10^{-10} \frac{\text{m}^2}{\text{s}}$ that it was also used previously in \cite{NoelJ1},\cite{NoelJ2}. Node S emits impulses of molecules to transmit a binary ``1'' with probability $P_{1}=0.5$. A binary sequence of length $L=50$ is transmitted, and the simulated bit error probability is averaged over $5 \times 10^{4}$ random realizations of the sequence. 
\begin{figure}[!t]
	\centering
	\includegraphics[scale=0.29]{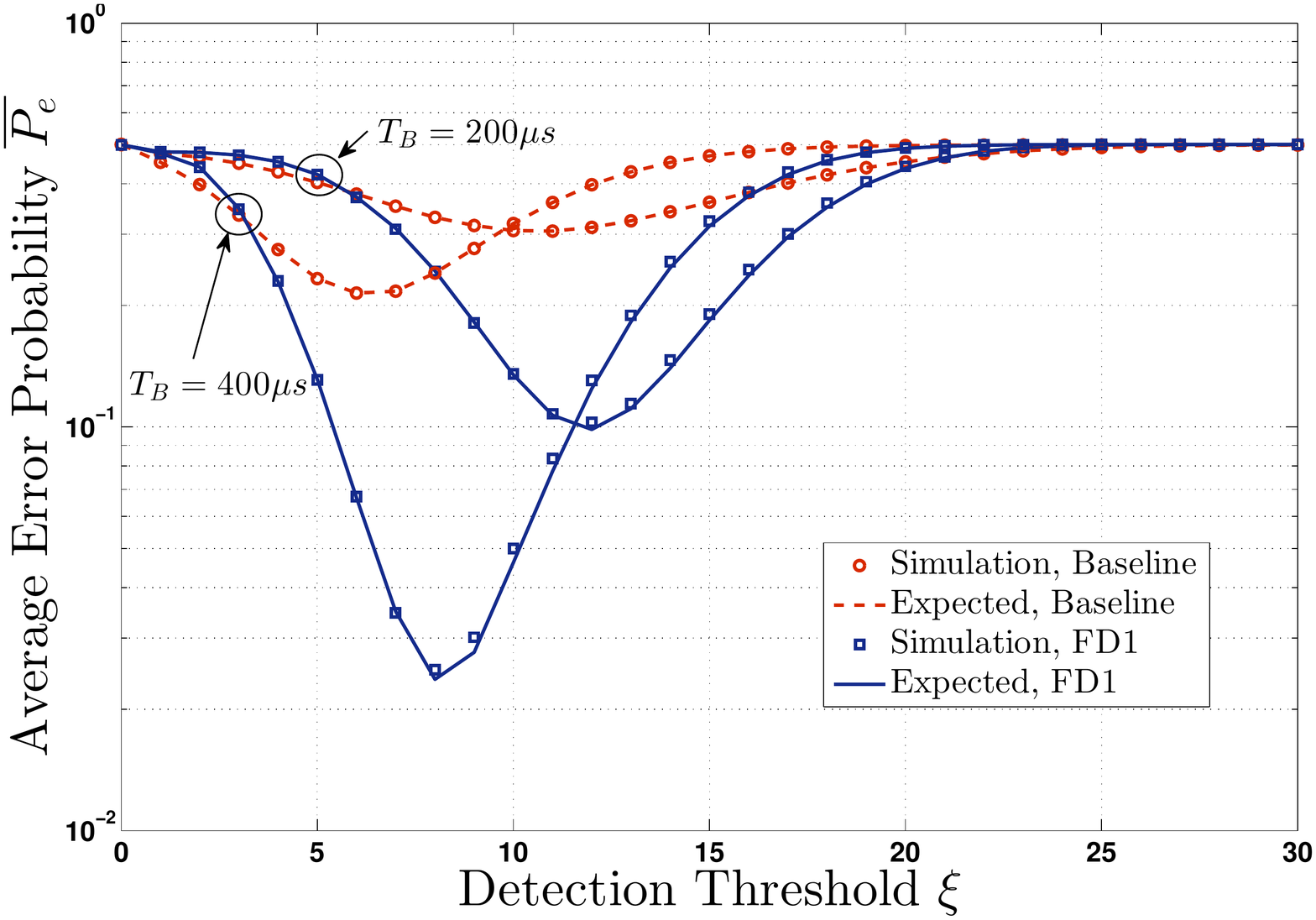}\vspace*{-3mm}
	\caption{Average error probability of a two-hop link as a function of the detection threshold, where different types of molecules are utilized in each hop.}
	\label{Fig.2}
\end{figure} 
\begin{figure}[!t]
	\centering
	\includegraphics[scale=0.29]{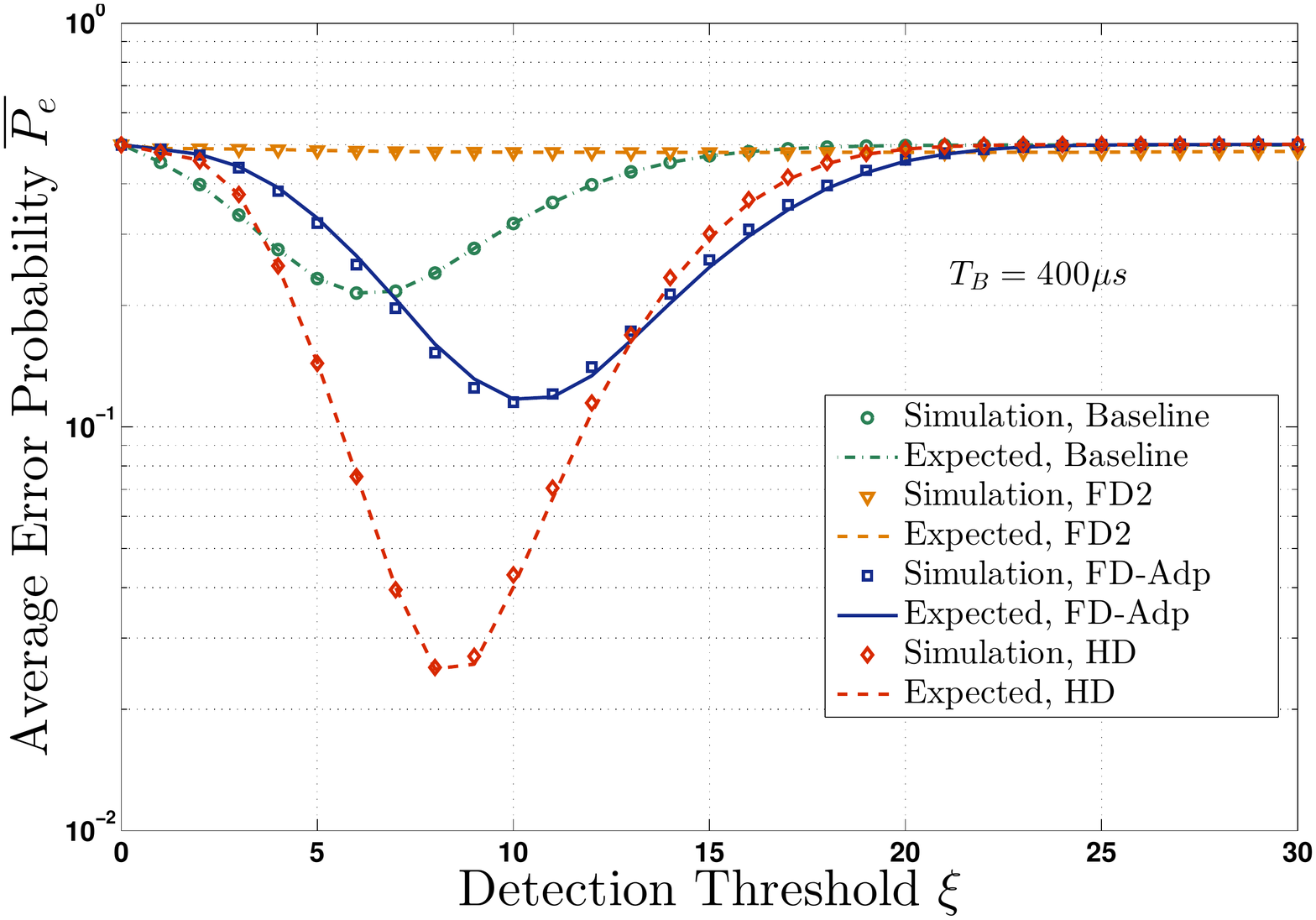}\vspace*{-3mm}
	\caption{Average error probability of a two-hop link as a function of the detection threshold, where the same type of molecules is utilized in each hop.}
	\label{Fig.3}
\end{figure}
\begin{table}
\renewcommand{\arraystretch}{1.3}
\caption{Summary of the Considered Relaying Protocols}
\label{Table1}
\centering
\begin{tabular}{|c|c|c|c|} 
\hline 
\bfseries Relaying & \bfseries Types of Molecules  & \bfseries Relay Detection & \bfseries Protocol \\
\bfseries Mode & \bfseries  in Both Hop & \bfseries Threshold & \bfseries Acronym \\ 
\hline 
Full-duplex & Different & Fixed & FD1 \\ 
\hline 
Full-duplex & Same & Fixed & FD2 \\ 
\hline 
Half-duplex & Same & Fixed & HD \\ 
\hline 
Full-duplex & Same & Adaptive & FD-Adp \\ 
\hline
\end{tabular}
\end{table} 
In Table \ref{Table1}, we summarize the protocols considered for two-hop transmission. The FD2 protocol, i.e., the full-duplex mode using the same type of molecule in each hop but without an adaptive threshold, is mainly considered for comparison to illustrate the effect of self-interference. 

In the following, we refer to the case when no relay is deployed between node S and node D as the \textit{baseline case}. The receiver is placed at $x_{D} = 600$ nm and $N_{A_{1}}=10000$ for the baseline case. For a fair comparison between the two-hop case and the baseline case, we assume that node R and node S release $5000$ molecules, respectively, to transmit information bit ``1'' for two-hop transmission. In all figures, the expected error probability of the two-hop link was evaluated via (\ref{Eq. ErrorExpSeq}), after taking into account the modifications required for each protocol.
 
In Figs. \ref{Fig.2} and \ref{Fig.3}, we assumed that the number of samples per bit interval is $M=5$ and $t_{0}=20$ $\mu$s. Fig. \ref{Fig.2} shows the performance of the two-hop link if different types of molecules are used in each hop as a function of the detection threshold, $\xiN{D}=\xiN{R}=\xi$, for $T_B = \{200,400\}$ $\mu$s. The results show that, by deploying one relay between the source and destination nodes, the overall performance of the network improves compared to the baseline case. Furthermore, by increasing $T_{B}$, the performance improves by an order of magnitude for the optimal detection threshold. This is because increasing $T_{B}$ decreases the effect of intersymbol interference (ISI). 

Fig. \ref{Fig.3} shows the average error probability of the two-hop link if the same type of molecules is used in each hop as a function of the detection threshold for $T_B = 400$ $\mu$s. For FD2, HD, and the baseline case, we adopted $\xiN{D}=\xiN{R}=\xi$, and for the FD-Adp protocol, the fixed part of the adaptive threshold is equal to $\xi$. Fig. \ref{Fig.3} shows that the FD2 protocol performs even worse than the baseline case. This confirms the performance limiting effect of self-interference. However, the proposed FD-Adp and HD protocols are effective in mitigating self-interference and perform better than the baseline scheme. Furthermore, the HD protocol performs better than the FD-Adp protocol. This is because for the FD-Adp protocol, the decision threshold can only be adapted based on the expected number of observed molecules, which may differ from the actual number of observed molecules. We note that the better performance of the HD protocol comes at the expense of decreasing the transmission rate by a factor of two. 

\begin{figure}[!t]
	\centering
	\includegraphics[scale=0.29]{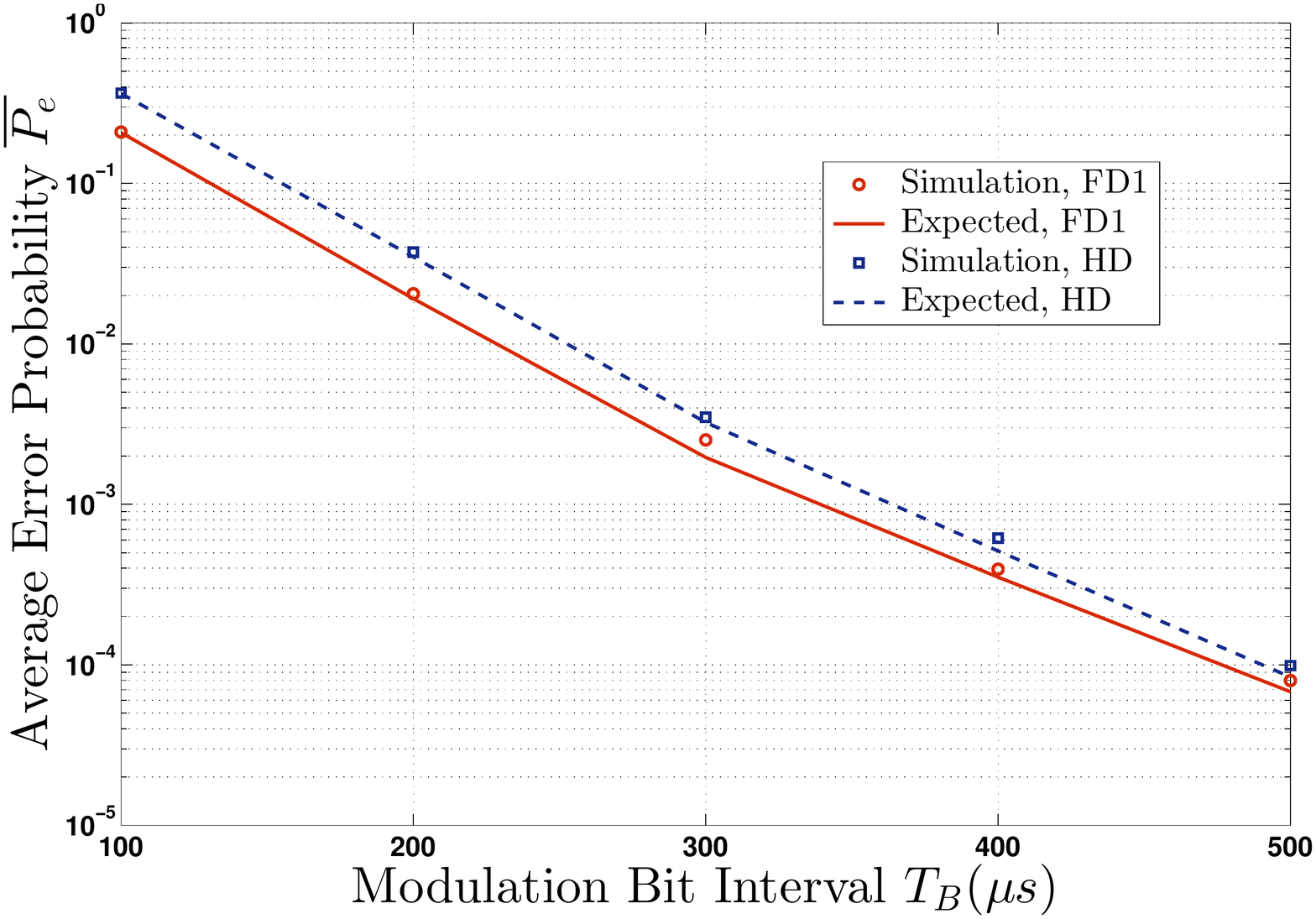}\vspace*{-3mm}
	\caption{Average error probability of a two-hop link as a function of modulation bit interval $T_B$ for the HD and FD1 protocols.}
	\label{Fig.4}
\end{figure} 
In Fig. \ref{Fig.4}, we compare the performance of the FD1 and HD protocols as a function of modulation bit interval $T_{B}$, for $M = 10$ and $t_{0} = 10$ $\mu$s. For each considered $T_B$, we use the optimal detection thresholds (which were found numerically) that minimize the expected error probabilities of the FD1 and HD protocols, respectively. The results in Fig. \ref{Fig.4} show that, by increasing the modulation bit interval, the HD protocol approaches the performance of the FD1  protocol. The reason for this is that, for large $T_{B}$, the molecules released by the relay for the HD protocol have enough time to leave the volume of the relay in one bit interval, and are not observed in the next bit interval when the relay takes samples for detection of the bit emitted by the source node.             
\section{CONCLUSION}
\label{Sec.Con} 
In this paper, we considered a two-hop link among nanomachines where we deployed one transceiver nanomachine between the transmitter and receiver nanomachines in an effort to improve the range of diffusion-based molecular communication. We considered two different relaying schemes. In the first scheme, different types of molecules were used in each hop and full-duplex relaying was adopted. In the second scheme, we used the same type of molecule in both hops, which led to self-interference at the relay. We considered two different approaches to mitigate the effect of self-interference: 1) An adaptive decision threshold at the relay, and 2) Half-duplex relaying instead of full-duplex relaying. We derived closed-form expressions for the expected error probability of two-hop transmission for all relaying schemes being considered. Our simulation and analytical results showed that the quality of communication between a transmitter nanomachine and a receiver nanomachine can be significantly improved by deploying a relay node. 

In our future work, we plan to extend the proposed framework to a general multihop network among nanomachines involving $\kappa$ relays.     

\bibliographystyle{IEEEtran}
\bibliography{Library}
\end{document}